\documentclass[pre,twocolumn,superscriptaddress,showpacs,aps]{revtex4-1}
\usepackage{graphicx}
\usepackage{amssymb, amsmath}
\usepackage{braket}
\usepackage{hyperref}
\hypersetup{colorlinks=true, linkcolor=blue, citecolor=blue, filecolor=blue, urlcolor=blue}

\begin{document}

\title{Signatures of coherent energy transfer and exciton delocalization in time-resolved optical cross correlations}

\author{Hallmann Óskar Gestsson}
\affiliation{Department of Physics and Astronomy, University College London, London WC1E 6BT, United Kingdom}
\author{Alexandra Olaya-Castro}
\email{a.olaya@ucl.ac.uk}
\affiliation{Department of Physics and Astronomy, University College London, London WC1E 6BT, United Kingdom}

\date{December 19, 2025}

\begin{abstract}
    We investigate how optical second-order cross correlations witness the quantum features of a prototype donor-acceptor light-harvesting unit. By considering a pair of detuned two-level emitters electronically coupled and incoherently driven to a non-equilibrium steady-state, we gain insight into how electronic quantum properties such as exciton eigenstate delocalization, coherent energy transfer and steady-state electronic coherence, are manifested in the joint probability of emission or optical second-order cross correlation. Specifically, we show that the frequency associated with oscillations present in time-resolved second-order cross correlation functions quantifies not only the time scale of coherent energy transfer but also the degree of delocalization of the exciton eigenstates. Furthermore, we show that time-resolved cross correlations directly witness steady-state electronic coherence. Our work strengthens the idea that measurements of the intensity quantum cross correlations can provide distinctive signatures of the quantum behavior of biophysical emitters. 
\end{abstract}

\maketitle

\section{Introduction}
Photosynthetic light-harvesting complexes are fascinating supramolecular units that contain chromophores bound to a protein scaffold. 
These complexes are responsible for the absorption of visible light and the subsequent primary excitation energy transfer process in photosynthetic organisms. 
Understanding and probing the quantum behavior of these complexes is a long standing scientific problem that has been the subject of intense scientific discussion over the last two decades \cite{Cheng2009May, Duan2017Aug, Kim2019May}.

It is well known that upon photoexcitation, electronic interactions between the chromophores in a light-harvesting complex lead to the formation of collective electronic states, or excitons, which are characterized by different degrees of delocalization across the complex \cite{vanAmerongen2000Jun}. 
This excitonic behavior is arguably the most fundamental and important quantum feature associated to such complexes. 
On the other hand, the possible quantum coherent nature of the excitation dynamics that follows photon absorption remains a subject of contentious scientific debate; despite remarkable experimental efforts \cite{Fleming2024Jan}, the presence and role of quantum coherence in photosynthetic complexes are yet to be unambiguously demonstrated \cite{Cao2020Apr, Kim2021Jan}. 
Therefore, an open problem in the field is to envision experimental probes that can provide insight into the signatures of different quantum features of excited state dynamics \cite{Fleming2024Jan}.

To verify the presence of non-trivial quantum effects in biomolecular structures, several proposals of spectroscopic witnesses have been put forward \cite{Smyth2012Aug, Marcus2019Dec, Otten2020Oct, SanchezMunoz2020May}. 
It has been theoretically demonstrated that the fluorescence properties of these systems can yield signatures of quantum coherence either when the system is coherently driven \cite{Marcus2019Dec, Otten2020Oct}, or incoherently \cite{delValle2010May, SanchezMunoz2020May, Downing2020Jul, Nation2024Feb}. 
Measurements of intensity auto correlations for biomolecular systems have been successfully carried out at the single molecule level, yielding clear photon anti-bunching, thereby demonstrating their nature as quantum emitters \cite{Berglund2002Jul, Hubner2003Aug}. 

In this work, we address the open problem of identifying signatures of quantum behavior of photo‑activated biophysical systems by investigating how time-resolved second-order cross correlations of emitted photons may report on quantum coherent dynamics, exciton delocalization and steady‑state electronic coherence.
We consider the simplest model system that can support coherent energy transfer, which is a pair of strongly interacting non-identical chromophores, namely, a heterodimer.
By focusing on the simplest scenario, we gain semi-analytical and physical insights that then can guide investigations into more complex scenarios. 
Previous theoretical studies have considered cross correlations for identical two-level emitters \cite{Downing2020Jul} and, therefore have not addressed the signatures associated with different degrees of exciton delocalization. 
Instead, our work is focused on the aspects uniquely associated with the detuning of the heterodimer. 
Experiments have demonstrated that sufficiently small separations between chromophores will enable coherent energy transfer to take place within the heterodimer as a result of their electronic coupling \cite{Kim2019May, Kong2022Jul}. 
Examples of strongly interacting pairs of chromophores can be found in several light-harvesting complexes, such as the nearest neighbor chromophores in the B850 ring in LH2 \cite{vanAmerongen2000Jun}, the central chromophore pair in PE545 \cite{Doust2004Nov} and in PC645 \cite{Marin2011Aug}, and the Venus$_\text{A206}$ dimer \cite{Kim2019May}. 
Furthermore, we consider the case of weak incoherent pumping to simulate the physiological conditions of the biomolecular systems of interest. In doing so, we demonstrate that quantum features that may be directly related to the biological function of the heterodimer: coherent energy transfer, exciton delocalization, and non-equilibrium steady-state coherence, all of which have the potential to be witnessed by time-resolved optical cross correlations. 

We provide semi-analytical expressions for the time-resolved cross correlations corresponding to emissions from a non-equilibrium steady-state. 
In particular, we show that such cross correlations are characterized by a frequency that is modulated by both the exciton energy gap and the exciton delocalization, and that the amplitude of such oscillations is determined by the degree of exciton delocalization. 
Furthermore, we consider a figure of merit for the time asymmetry present in the cross correlations and show that its magnitude is positively correlated with the steady-state electronic coherence. 
Therefore, our work demonstrates the potential of time-resolved cross correlations to report on the non-trivial quantum aspects of composite emitter complexes.

The remainder of this paper is organized as follows. In Section \ref{Sec:Hamiltonian}, we introduce the Hamiltonian for the heterodimer system and the quantum master equation required to describe the incoherent processes. We also present analytical expressions for the steady-state populations and coherences.
In Section \ref{Sec:PopulationDynamics}, we present the equations of motion for the populations and coherences relevant for the computation of time-resolved cross correlations. 
In Section \ref{Sec:2ndOrderCorrelations}, we introduce the time-resolved second-order cross correlation function and derive semi-analytical expressions for the case of interest. 
We then relate the time asymmetry of the cross correlation to steady-state electronic coherence. 
The conclusions and outlook are presented in Section \ref{sec:conclusions}. 

\begin{figure}
    \centering
    \includegraphics[width=\linewidth]{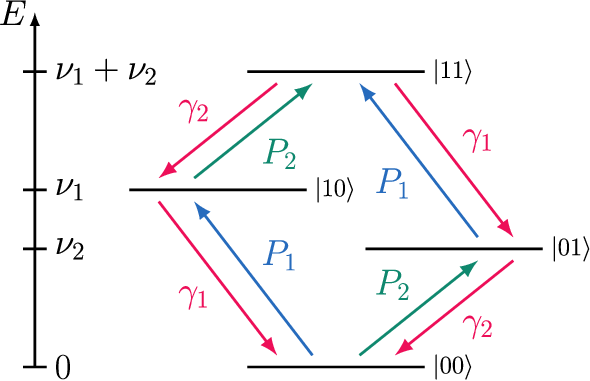}
    \caption{Level structure of the site basis states. One-way arrows indicate incoherent pathways between the site states of the system that arise due to radiative decay and incoherent pumping of the electronic states.}
    \label{fig:system_structure}
\end{figure}

\section{Dissipative electronic system and its steady-state}\label{Sec:Hamiltonian}
\subsection{Heterodimer system description and exciton features}\label{Sec:DimerDescription}
We consider a pair of electronically coupled chromophores and approximate each as a two-level emitter that is characterized by their respective transition frequency $\nu_i$, where $i=1,2$ indexes the chromophores. 
The local Hamiltonian of each chromophore is then of the form $\hat{H}_i = \hbar\nu_i\hat{n}_i$ with emitter population operators $\hat{n}_i = \sigma_i^+\sigma_i^-$ where $\sigma_i^+$ ($\sigma_i^-$) is the raising (lowering) operator of the $i$-th emitter. 
The state space of $\hat{H}_i$ is spanned by a basis $\mathcal{B}_i = \left\{\ket{0}_i,\ket{1}_i\right\}$ in which $\ket{0}_i$ and $\ket{1}_i = \sigma_i^+\ket{0}_i$ represent the ground and excited state, respectively. 
Furthermore, the pair is electronically coupled via a transition dipole interaction term $\hat{H}_{int} = \hbar V(\sigma_1^+\sigma_2^- + \text{h.c.})$ where for simplicity, the coupling strength $V$ is considered to be real-valued. 
The total Hamiltonian for our prototypical heterodimer system is then given as ($\hbar = 1$ throughout the remainder of this article)
\begin{equation}
    \hat{H} = \nu_1\hat{n}_1 + \nu_2\hat{n}_2 + V(\sigma_1^+\sigma_2^- + \text{h.c.})
\end{equation}
The state space of $\hat{H}$ is spanned by a basis given by the tensor product of the two local basis $\mathcal{B} = \left\{\ket{00},\ket{10},\ket{01},\ket{11}\right\}$, where we've adopted the notation $\ket{k}_1\otimes\ket{l}_2 \equiv \ket{kl}$. 
Throughout this paper, we will refer to $\mathcal{B}$ and its elements as the \textit{site basis states}. 
Fig.\ \ref{fig:system_structure} illustrates the energy level structure of the site basis states and the incoherent pathways induced by an environment interacting with the dimer system, whose description is presented in Section \ref{Sec:BMME}. 
Note that the eigenstates of $\hat{H}$ coincide only with the site states when the coupling strength is zero. Non-zero coupling strengths induce the formation of \textit{exciton} eigenstates within the single-excitation manifold of $\hat{H}$. 
The energy eigenvalues of these excitons are $E_\pm = \frac{\nu_1+\nu_2}{2} \pm \frac{1}{2}\Delta E$, where $\Delta E$ is the exciton energy gap,
\begin{equation}
    \Delta E = \sqrt{4V^2+\delta^2},
\end{equation}
with $\delta = \nu_1 - \nu_2$ being the energy offset between the two bare single excitation site states, $\ket{10}$ and $\ket{01}$. The timescale of coherent energy transfer between chromophores is determined by $\Delta E$.

\begin{figure}
    \centering
    \includegraphics[width=\linewidth]{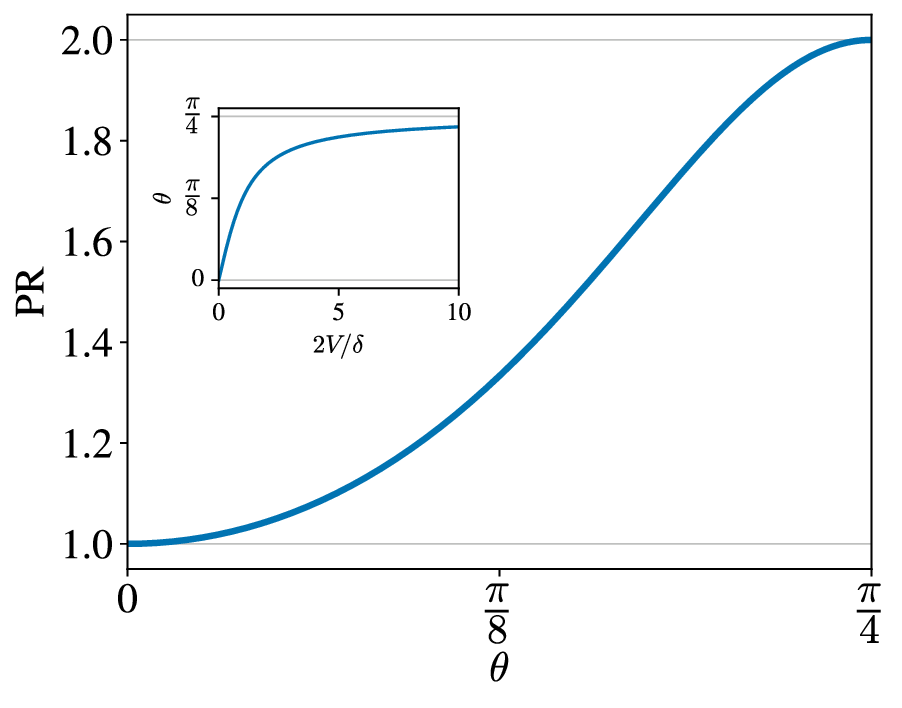}
    \caption{Participation ratio as a function of the mixing angle. The inset shows the mixing angle as a function of the ratio $2V/\delta$, as determined by Eq.\ (\ref{eq:mixing_angle}). For $\theta = 0$ the excitons will be completely localized to a single site and for $\theta = \pm\frac{\pi}{4}$ the excitons will be completely delocalized across both sites.}
    \label{fig:participation_ratio}
\end{figure}

The exciton states are a linear combination of the site states, which can be expressed as
\begin{align}\label{eq:excitons}
    \ket{-} & = \cos\theta\ket{10} - \sin\theta\ket{01}, \\
    \ket{+} & = \sin\theta\ket{10} + \cos\theta\ket{01},
\end{align}
where we have introduced the \textit{mixing angle}
\begin{equation}\label{eq:mixing_angle}
    \theta = \frac{1}{2}\arctan\left(\frac{2V}{\delta}\right),
\end{equation}
which indicates the degree to which excitons are delocalized across the heterodimer. Using the mixing angle as a figure of merit for delocalization may be justified by considering its relation to the participation ratio, which in our case becomes $PR = (\cos^4\theta + \sin^4\theta)^{-1}$ \cite{Thouless1974Oct, Scholes2014Mar}. 
Fig.\ \ref{fig:participation_ratio} illustrates the participation ratio as a function of the mixing angle. Note that for mixing angles tending towards $\pm\frac{\pi}{4}$ corresponds to the case where $\vert V\vert \gg \delta$ such that the excitons are completely delocalized across the dimer whilst for small mixing angles $\theta\approx 0$ the excitons become highly localized to their respective site state. 
Another motivation for introducing the mixing angle is that it allows for the following parameterization of the system parameters,
\begin{align}
    \label{eq:parameterisation1}
    V &= \frac{1}{2}\Delta E\sin(2\theta),\\ \label{eq:parameterisation2}
    \delta &= \Delta E\cos(2\theta),
\end{align}
such that $\Delta E$ may be kept fixed for different degrees of exciton delocalization. 

\subsection{Born-Markov master equation for radiative decay and incoherent pumping}\label{Sec:BMME}
To study the optical response of our prototypical model, we apply a typical Lindblad master equation approach to obtain a description of the heterodimer system dynamics as they are influenced by incoherent radiative processes that stem from a weak interaction with a reservoir \cite{BRE02}. 
For each incoherent process, we associate a Lindbladian superoperator $\mathcal{L}_O$ whose action on the reduced density matrix $\rho$ is given as $\mathcal{L}_O(\rho) = 2O\rho O^\dagger - O^\dagger O\rho - \rho O^\dagger O$, where $O$ is an operator that controls the effect of the incoherent process on $\rho$. 

The dynamics of the reduced density matrix for the electronically coupled heterodimer system are determined by the quantum master equation
\begin{equation}\label{Eq:QuantumMasterEquation}
    \partial_t\rho = -i\left[\hat{H},\rho\right] + \frac{1}{2}\sum_{i=1,2}\left(\gamma_i\mathcal{L}_{\sigma_i^-}(\rho) + P_i\mathcal{L}_{\sigma_i^+}(\rho)\right),
\end{equation}
where unitary dynamics are described by the commutator term and incoherent effects are induced by the $\mathcal{L}_O$ superoperators. 
Each emitter is assumed to radiatively decay at a rate $\gamma_i$ while simultaneously being incoherently pumped at a rate $P_i$. 
These pathways between the site basis states are illustrated in Fig.\ \ref{fig:system_structure} as one-way arrows to reflect their incoherent nature. 
Eq.\ (\ref{Eq:QuantumMasterEquation}) and its steady-state has been studied in the literature before, see e.g. \cite{SanchezMunoz2020May,delValle2010May,Downing2020Jul}. 

When analyzing the steady-state and transient dynamics of populations, it will be convenient to work in terms of combinations of incoherent radiative rates that are even and odd with respect to a swapping of indexes, which we denote as
\begin{align}
    \gamma_e &= \frac{\gamma_1 + \gamma_2}{2} + \frac{P_1 + P_2}{2}, \\
    \gamma_o &= \frac{\gamma_1 - \gamma_2}{2} + \frac{P_1 - P_2}{2},
\end{align}
where $\gamma_e$ and $\gamma_o$ are even and odd combinations, respectively. The rate at which the time-resolved second-order cross correlation decays is given by $\gamma_e$, which is the average decay rate as a result of every incoherent process acting on the system. On the other hand, $\gamma_o$ is a more subtle quantity that may act as a figure of merit for the imbalance of incoherent population transfer.
As we shall see in Section \ref{Sec:2ndOrderCorrelations}, a vanishing $\gamma_o$ will result in a second-order cross correlation function that is symmetric in time. For the remainder of this paper we assume identical decay rates, that is, $\gamma_1 = \gamma_2 = \gamma$, such that $\gamma_e = \gamma + \frac{P_1 + P_2}{2}$ and $\gamma_o = \frac{P_1 - P_2}{2}$. 

\subsection{Steady-state populations and coherences}\label{Sec:SteadyStateElements}
\begin{figure}
    \centering
    \includegraphics[width=\linewidth]{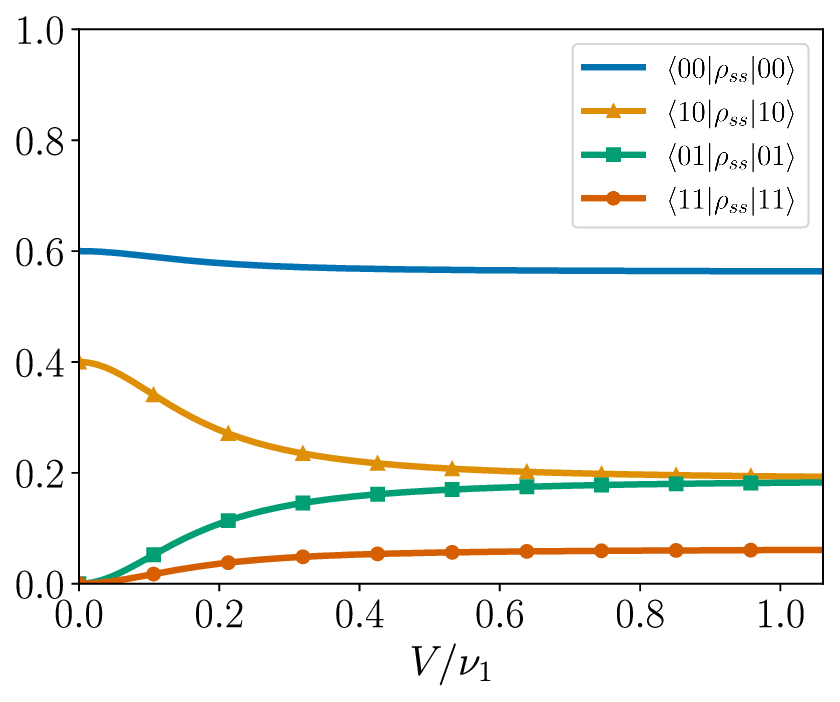}
    \caption{steady-state populations in the site basis as a function of the electronic coupling strength $V$ scaled by the transition frequency $\nu_1$. Here we have set $\delta = \gamma_e = 0.25$, $P_2 = 0$ and $\frac{P_1}{2\gamma} = \frac{1}{3}$. For uncoupled emitters we have the steady-state populations to be completely determined by the radiative rates. As the coupling strength increases, the second site becomes populated leading to non-zero electronic coherence in the steady-state (see Eq.\ (\ref{eq:SteadyStateCoherence})). In the strong coupling limit, i.e.\ $V\rightarrow\infty$, site populations reach the same value thereby leading to a vanishing steady-state coherence. }
    \label{Fig:SteadyPopulations}
\end{figure}

Every initial state of the system will eventually relax into a steady-state, which is determined by the condition $\partial_t\rho_{ss} = 0$, where $\rho_{ss}$ is the steady-state density matrix. 
In terms of the site state basis, we have according to Eq.\ (\ref{Eq:QuantumMasterEquation}) that every site population and the single excitation coherences are coupled, forming a closed system of equations involving six variables: four populations, and the conjugate pair of single excitation coherences. 
All the other coherence terms are uncoupled and decay exponentially from their initial condition to zero. 
We can express the steady-state populations in the form
\begin{align}
    \braket{00\vert\rho_{ss}\vert 00} &= \left[\frac{4P_1P_2 + (P_1 + P_2)^2\kappa^2}{4\gamma^2(1+\kappa^2)} + \frac{\left(P_1 + P_2\right)}{\gamma} + 1\right]^{-1}, \label{Eq:SteadyStatePopulation1}\\
    \braket{10\vert\rho_{ss}\vert 10} &= \frac{2P_1 + (P_1 + P_2)\kappa^2}{2\gamma(1+\kappa^2)}\braket{00\vert\rho_{ss}\vert 00}, \label{Eq:SteadyStatePopulation2}\\
    \braket{01\vert\rho_{ss}\vert 01} &= \frac{2P_2 + (P_1 + P_2)\kappa^2}{2\gamma(1+\kappa^2)}\braket{00\vert\rho_{ss}\vert 00}, \label{Eq:SteadyStatePopulation3}\\
    \braket{11\vert\rho_{ss}\vert 11} &= \frac{4P_1P_2 + (P_1 + P_2)^2\kappa^2}{4\gamma^2(1+\kappa^2)}\braket{00\vert\rho_{ss}\vert 00}, \label{Eq:SteadyStatePopulation4}
\end{align}
where we introduced the ratio $\kappa = \frac{2V}{\sqrt{\gamma_e^2 + \delta^2}}$ to simplify the notation. See Appendix \ref{Appendix:SteadyStatePopulations} for further details on how to derive these expressions. 
Fig.\ \ref{Fig:SteadyPopulations} shows the steady-state populations as a function of the coupling strength $V$ with $P_1 = \frac{2}{3}\gamma$ and $P_2 = 0$, which demonstrates how the populations of single excitation states will tend towards one another for very large values of $V$. 
For a moderate electronic coupling strength, the populations of single‑excitation states will be different and will lead to non‑zero steady‑state electronic coherences as can be concluded from the equation below:
\begin{equation}\label{eq:SteadyStateCoherence}
    \braket{10\vert\rho_{ss}\vert 01} = \frac{\kappa}{2}\frac{\delta + i\gamma_e}{\sqrt{\delta^2 + \gamma_e^2}}\left(\braket{10\vert\rho_{ss}\vert 10} - \braket{01\vert\rho_{ss}\vert 01}\right),
\end{equation}
which has been analyzed as a signature of quantum effects within dimer systems \cite{SanchezMunoz2020May}. 

For balanced pumping, that is, $P_1 = P_2 = P$, the site populations become identical such that steady-state coherences vanish, and steady-state site populations depend only on the incoherent rates $\gamma$ and $P$. We will have
\begin{align}\label{Eq:BalancedSteadyStatePopulations}
    \braket{00\vert\rho_{ss}\vert 00} &= \frac{\gamma^2}{\left(P + \gamma\right)^2}, \\
    \braket{10\vert\rho_{ss}\vert 10} &= \braket{01\vert\rho_{ss}\vert 01} = \frac{P\gamma}{\left(P + \gamma\right)^2}, \\ \braket{11\vert\rho_{ss}\vert 11} &= \frac{P^2}{\left(P + \gamma\right)^2}.
\end{align}
Note that $\gamma_o = 0$ results in an equilibrium steady-state, despite the fact that the system may be composed of two non-identical emitters. 
In Section \ref{Sec:2ndOrderCorrelations}, we see that this condition results in time-resolved second-order cross correlations that are symmetric in time. 

\section{Dynamics of populations and coherences}\label{Sec:PopulationDynamics}
As we will show below, the dynamical evolution of populations in the site basis can be related to the time-resolved second-order correlation function by virtue of the quantum regression theorem. 
Therefore, we gain analytical insight into the correlation functions by analyzing population dynamics. 
To do so, we make use of the fact that the mean value dynamics for an arbitrary Heisenberg picture system operator $\hat{A}$ is determined by $\partial_t\braket{\hat{A}} = \text{Tr}(\hat{A}(0)\partial_t\rho)$, where $\hat{A}(0)$ is the corresponding Schrödinger picture operator and $\partial_t\rho$ is given by Eq.\ (\ref{Eq:QuantumMasterEquation}). For $\hat{A} = \hat{n}_1$ we obtain the following equation of motion,
\begin{equation}
    \partial_t\braket{\hat{n}_1} = P_1 - (\gamma+P_1)\braket{\hat{n}_1} -iV(\braket{\sigma_1^+\sigma_2^-} - \braket{\sigma_1^-\sigma_2^+}),
\end{equation}
where we use the cyclic property of the trace operation. The set of operators $\left\{\hat{n}_1,\hat{n}_2,\sigma_1^+\sigma_2^-,\sigma_1^-\sigma_2^+\right\}$ form a closed linear system of differential equations. Defining the column vector $\vec{x}(t) = \left(\braket{\hat{n}_1},\braket{\hat{n}_2},\braket{\sigma_1^+\sigma_2^-},\braket{\sigma_1^-\sigma_2^+}\right)^T$ we obtain the matrix differential equation
\begin{equation}\label{Eq:AdjointMasterEq}
    \partial_t\vec{x}(t) = M\vec{x}(t) + \vec{b}
\end{equation}
where $\vec{b} = \left(P_1,P_2,0,0\right)^T$ and $M$ is the matrix
\begin{equation}
    M = \begin{bmatrix}
    -\gamma_e - \gamma_o & 0 & -iV & iV \\
    0 & -\gamma_e + \gamma_o & iV & -iV \\
    -iV & iV & -\gamma_e + i\delta & 0 \\
    iV & -iV & 0 & -\gamma_e - i\delta
    \end{bmatrix},
\end{equation}
which we have written in terms of $\gamma_e$ and $\gamma_o$. This system of equations for the system's observables arises in similar contexts \cite{Laussy2009Jun,delValle2010May}. The formal solution of the matrix equation is $\vec{x}(t) = \vec{x}_{ss} + e^{Mt}\left(\vec{x}_0 - \vec{x}_{ss}\right)$, where $\vec{x}_0$ is the initial condition and $\vec{x}_{ss} = -M^{-1}\vec{b}$ is the steady-state solution towards which our system relaxes. The dynamical propagator $e^{Mt}$ determines the temporal evolution of $\vec{x}$ (and, therefore, $\braket{\hat{n}_1}$ and $\braket{\hat{n}_2}$). 

\begin{figure}
    \centering
    \includegraphics[width=\linewidth]{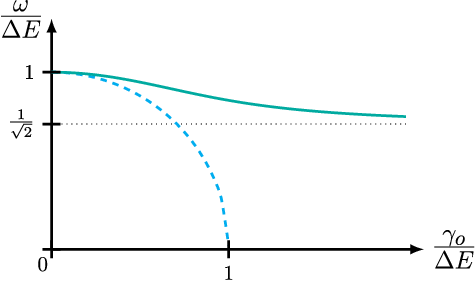}
    \caption{Graph of $\omega/\Delta E$ as a function of $\gamma_o/\Delta E$ as determined by Eq.\ (\ref{Eq:MainFreq}). The dashed line corresponds to the case $\theta=\frac{\pi}{4}$ whilst the solid line corresponds to $\theta = \frac{\pi}{8}$. The dotted line is the limit of $\omega$ when $\gamma_o$ tends towards infinity and is determined as $\frac{\delta}{\Delta E}$.}
    \label{Fig:FreqPumping}
\end{figure}

Eigenvalues of $M$ allow us to deduce the characteristic features of the population dynamics, namely that a negative real component corresponds to a decay rate and an imaginary component corresponds to oscillatory behavior. The eigenvalues of $M$ are of the form
\begin{align}
    \lambda_1 &= -\gamma_e-i\omega, \\ 
    \lambda_2 &= -\gamma_e+i\omega, \\
    \lambda_3 &= -\gamma_e-\frac{\delta}{\omega}\gamma_o, \\
    \lambda_4 &= -\gamma_e+\frac{\delta}{\omega}\gamma_o,
\end{align}
where $\omega$ is defined as
\begin{equation}\label{Eq:MainFreq}
    \omega = \frac{\Delta E}{\sqrt{2}}\sqrt{1 - \frac{\gamma_o^2}{\Delta E^2} + \sqrt{\left(1 - \frac{\gamma_o^2}{\Delta E^2}\right)^2 + 4\frac{\delta^2}{\Delta E^2}\frac{\gamma_o^2}{\Delta E^2}}}.
\end{equation}
As the evolution of populations and coherences are governed by a linear combination of exponential functions whose decay rates are determined by the eigenvalues of $M$, any oscillations observed must have their frequency determined by Eq.\ \eqref{Eq:MainFreq}. Recalling the parametrization given in Eqs.\ (\ref{eq:parameterisation1}-\ref{eq:parameterisation2}), it becomes clear that $\omega$ is explicitly a function of the mixing angle $\theta$ when pumping is imbalanced, that is $\gamma_o\neq0$. This means that population oscillations are able to directly witness exciton delocalization and provide a means to quantify the mixing angle in an experiment. Fig.\ \ref{Fig:FreqPumping} shows $\omega/\Delta E$ as a function of the scaled pumping imbalance $\gamma_o/\Delta E$ for two cases of mixing angles, $\theta = \pi/4$ and $\theta = \pi/8$. There is a clear qualitative difference between the two cases stemming from the fact that for completely delocalized excitons ($\theta = \pi/4$) in the regime where $\gamma_o \geq \Delta E$ we have the population oscillations to be completely suppressed and exponentially damped towards their steady-state. For quasi-localized excitons ($\theta = \pi/8$) we find that $\omega$ tends towards $\delta$ when $\vert\gamma_0\vert$ tends towards infinity. However, this does not imply that we might witness oscillations in the $\vert\gamma_o\vert\rightarrow\infty$ limit because they will be effectively damped due to the large average damping rate $\gamma_e$.

\begin{figure}
    \centering
    \includegraphics[width=\linewidth]{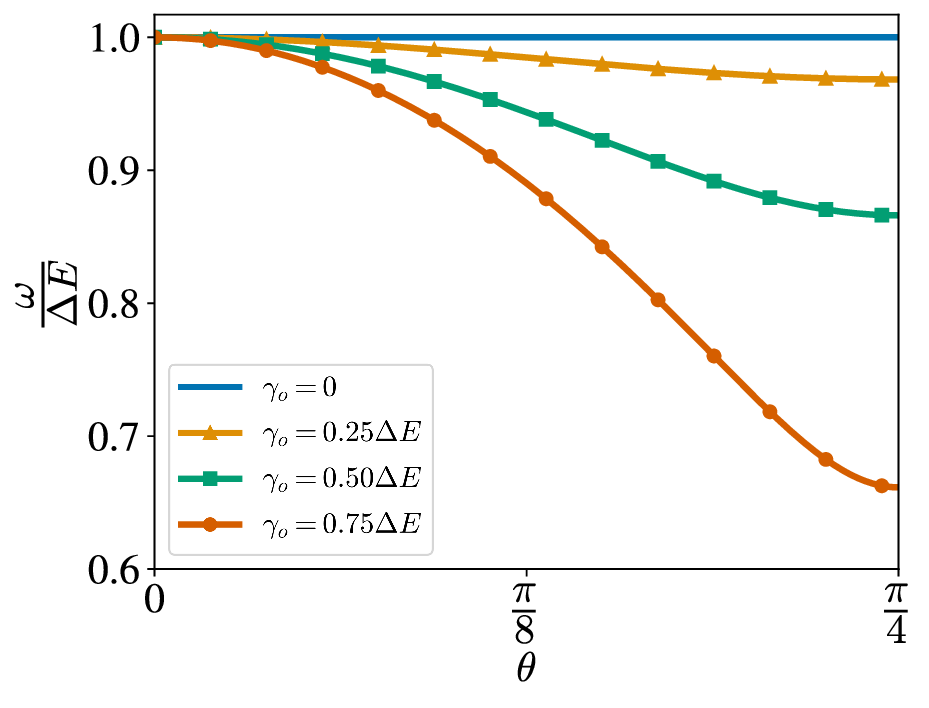}
    \caption{The figure shows $\omega$ scaled by the exciton energy gap as a function of the mixing angle $\theta$. The lines represent the deviation of $\omega$ away from $\Delta E$ as a function of $\theta$ for several values of $\frac{\gamma_o}{\Delta E}$, as indicated by the legend. }
    \label{Fig:freq_dep_v_g_o}
\end{figure}

In a regime of weak pumping where $\gamma_o/\Delta E \ll 1$ we have that a second-order Taylor expansion of Eq.\ (\ref{Eq:MainFreq}) in terms of $\gamma_o/\Delta E$ yields
\begin{equation}\label{Eq:MainTaylor}
    \omega = \Delta E\left(1 - \frac{1}{2}\sin^2(2\theta)\frac{\gamma_o^2}{\Delta E^2}\right) + \mathcal{O}\left(\frac{\gamma_o^4}{\Delta E^4}\right).
\end{equation}
Note that we make no assumption regarding the mixing angle such that Eq.\ (\ref{Eq:MainTaylor}) is a good approximation for all possible values of $\theta$ as long as $\gamma_o/\Delta E \ll 1$. From the expansion, we see that exciton delocalization will manifest as a deviation in $\omega$ away from the case of no delocalization, where the expansion would yield $\Delta E  - \frac{\gamma_o^2}{2\Delta E}$. Therefore, the frequency of oscillations that populations may exhibit will provide us with a direct signature of exciton delocalization, and the imbalanced pumping of the heterodimer is a crucial component if one is to witness this effect. This is shown in Fig.\ \ref{Fig:freq_dep_v_g_o}, which plots the change in $\omega$ with increasing delocalization, as determined by Eq.\ (\ref{Eq:MainFreq}) for select values of $\gamma_o$. We see that in the case where the pumping is symmetric ($\gamma_o = 0$), $\omega$ will not be modified by exciton delocalization, demonstrating the importance of imbalanced pumping to witness exciton delocalization. 

\begin{figure*}
    \centering
    \includegraphics[width=\textwidth]{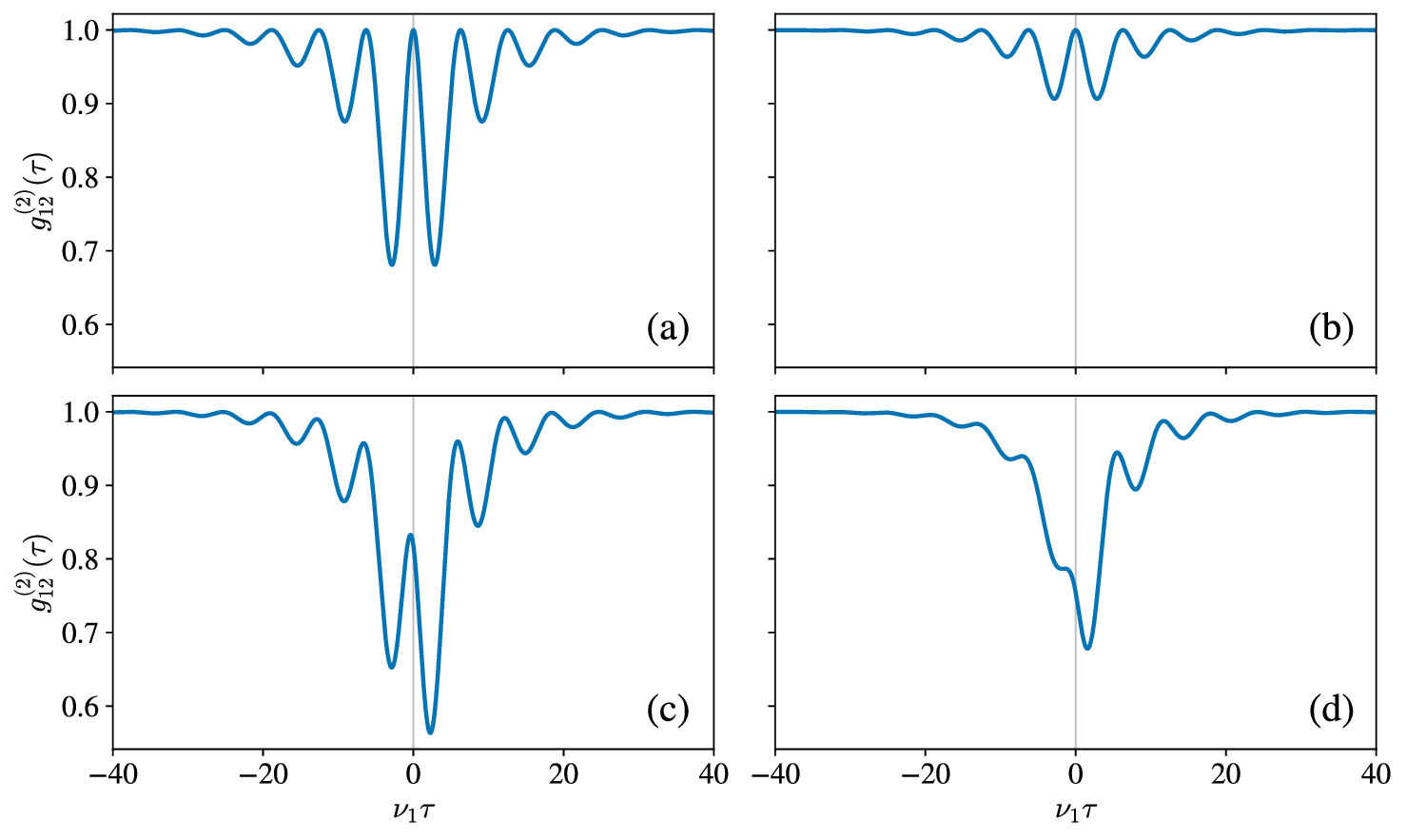}
    \caption{Time-resolved second-order cross correlation functions for the electronically coupled dimer system with radiative decay rates $\gamma = 0.1\nu_1$. The plots have identical $x$- and $y$-axes, and we enforce $\gamma_e$ and $\Delta E$ to be constant for every case. For (a) and (b) there is balanced pumping with $P_1 = P_2 = 0.05\nu_1$ which results in symmetric correlations in both cases. We have $\theta = \pi/8$ and $\theta = \pi/16$ for (a) and (b), respectively, illustrating the scaling in amplitude of oscillations for different degrees of localization. For (c) and (d) there is imbalanced pumping with $P_1 = 0.099\nu_1$ and $P_2 = 0.001\nu_1$ which results in asymmetric correlations in both cases. We have $\theta = \pi/8$ and $\theta = \pi/16$ for (c) and (d), respectively.}
    \label{Fig:g2tau_subplots}
\end{figure*}

\section{Optical time-resolved second-order cross correlations}\label{Sec:2ndOrderCorrelations}
Second-order correlation functions describe the emission coincidence statistics of photons radiated by a quantum system. 
Specifically, time-resolved correlation functions describe correlations between two successively radiated photons, in which it is assumed that the first emission event occurs at a fixed time $t$ and then the second emission event is assumed to occur after a delay, which we denote $\tau$. 
The $\tau = 0$ case then corresponds to the correlations between two photons emitted simultaneously.
Understanding how the quantum mechanical nature of the driven-dissipative system manifests in the second-order correlations provides insight into how a measurement of time-resolved photon emission coincidences may witness both coherent energy transfer and exciton delocalization. 

The heterodimer model we consider consists of two electronically coupled quantum emitters, and we consider the normalized second-order cross correlations for these two emitters,
\begin{equation}
    g_{12}^{(2)}(t,t+\tau) = \frac{\braket{\sigma_1^+(t)\hat{n}_2(t+\tau)\sigma_1^-(t)}}{\braket{\hat{n}_1(t)}\braket{\hat{n}_2(t+\tau)}},
\end{equation}
where we have assumed a positive time delay $\tau\geq0$. 
For $\tau < 0$ indices are swapped. 
The cross correlation function describes the joint probability of emissions, that is, the probability of a second emission occurring from emitter $2$ after a delay $\tau$ given that an emission from emitter $1$ occurred at an earlier time $t$ \cite{scully_zubairy_1997}. 
After normalizing the correlation function by uncorrelated emissions, $\braket{\hat{n}_1(t)}\braket{\hat{n}_2(t+\tau)}$, we have $g_{12}^{(2)}(t,t+\tau) = 1$ corresponding to uncorrelated photon emissions. 
As the chromophores are electronically coupled to one another, they will not behave as independent emitters, and from the features of the correlations we can deduce information regarding internal system dynamics. 
The exponential decay of the correlations is associated with the presence of incoherent processes, and oscillatory behavior can be associated with both coherent energy transfer within the heterodimer and exciton delocalization. 

Furthermore, we assume that emissions occur from the steady-state of the system, meaning that the timescale for system relaxation is much shorter than the timescale of radiative decay. 
This corresponds to taking the $t\to\infty$ limit,
\begin{equation}\label{Eq:g2SteadyStateLimit}
    g^{(2)}_{12}(\tau) = \lim_{t\to\infty}g_{12}^{(2)}(t,t+\tau) = \frac{\braket{\sigma^+_1(0)\hat{n}_2(\tau)\sigma^-_1(0)}_{ss}}{\braket{\hat{n}_1}_{ss}\braket{\hat{n}_2}_{ss}}
\end{equation}
where the subscript $ss$ indicates that the trace is taken with respect to the steady-state density matrix $\rho_{ss}$ which was discussed in Section \ref{Sec:SteadyStateElements}. 
Having determined the matrix coefficients of $\rho_{ss}$ with respect to the site basis $\mathcal{B}$ we are able to rewrite Eq.\ (\ref{Eq:g2SteadyStateLimit}) in terms of the steady-state site populations and the dynamics of the site populations with specified initial conditions. 
The result is an expression that is piece-wise due to the time-ordering of the correlation function,
\begin{widetext}
    \begin{equation}\label{Eq:g2tauPopulations}
        g^{(2)}_{12}(\tau) = \frac{1}{\braket{\hat{n}_1}_{ss}\braket{\hat{n}_2}_{ss}}
            \begin{cases}
           \braket{10\vert\rho_{ss}\vert 10}\braket{\hat{n}_2(\tau)}_{\rho_0=\ket{00}\bra{00}} + \braket{11\vert\rho_{ss}\vert 11}\braket{\hat{n}_2(\tau)}_{\rho_0=\ket{01}\bra{01}},& \tau\geq 0, \\[5pt]
           \braket{01\vert\rho_{ss}\vert 01}\braket{\hat{n}_1(\tau)}_{\rho_0=\ket{00}\bra{00}} + \braket{11\vert\rho_{ss}\vert 11}\braket{\hat{n}_1(\tau)}_{\rho_0=\ket{10}\bra{10}},& \tau< 0,
        \end{cases}
    \end{equation}
\end{widetext}
where the $\rho_0$ subscript indicates the initial density matrix of the system. 
Eq.\ (\ref{Eq:g2tauPopulations}) directly relates the emitter population dynamics to time-resolved cross correlations and therefore demonstrates that Rabi oscillations manifest in the correlations. 
Let us now consider the $\tau\geq 0$ case, where we have two terms, each with their respective weighting that corresponds to the probability that emitter 1 dissipates a photon from either the single or double excitation manifold. 
Therefore, the two terms represent the ensuing dynamics that stem from the initial state in which either event will condition the system to. 

Eq.\ (\ref{Eq:g2tauPopulations}) in terms of elementary functions is of the form
\begin{equation}\label{Eq:g2tauelementary}
    \begin{aligned}
        g^{(2)}_{12}(\tau) = 1 - e^{-\gamma_e\vert\tau\vert}\bigg(&A_\pm\cos(\omega\tau) + B_\pm\sin(\omega\tau) \\
        &+ C_\pm e^{\frac{\delta}{\omega}\gamma_o\tau} + D_\pm e^{-\frac{\delta}{\omega}\gamma_o\tau}\bigg),
    \end{aligned}
\end{equation}
where $A_\pm$, $B_\pm$, $C_\pm$, $D_\pm$ are constants with a $\pm$ subscript indicating the sign of $\tau$. The constants have cumbersome expressions in terms of the system parameters $V$, $\delta$, and the incoherent rates $\gamma$, $P_1$, $P_2$, but Eq.\ (\ref{Eq:g2tauelementary}) is nonetheless useful to consider to obtain an idea of the qualitative behavior of the second-order cross correlations and to elucidate the different roles that $\gamma_e$ and $\gamma_o$ serve in the correlation function. As noted in Section \ref{Sec:BMME}, $\gamma_e$ acts as the average decay rate of the cross correlations, which determines the timescale in which emissions eventually become uncorrelated. Note that we have terms that decay at rates $\gamma_e+\frac{\delta}{\omega}\gamma_o$ and $\gamma_e-\frac{\delta}{\omega}\gamma_o$ but $\gamma_e$ nonetheless is the average decay rate of the second-order cross correlations. Meanwhile, the pumping imbalance $\gamma_o$ modifies $\omega$, as per Eq.\ (\ref{Eq:MainFreq}).

When the pumping of both emitters is balanced ($P_1 = P_2$) we find that Eq.\ (\ref{Eq:g2tauelementary}) becomes symmetric with respect to $\tau$ and is reduced to a simple form. In Section \ref{Sec:SteadyStateElements}, it was established that the condition $\gamma_o = 0$ will result in the single excitation steady-states to be equal to one another, $\braket{10\vert\rho_{ss}\vert 10} = \braket{01\vert\rho_{ss}\vert 01}$, and in Appendix \ref{Appendix:DynamicsAppendix} it is determined as well that $\gamma_o = 0$ will imply that $\braket{\hat{n}_1(\tau)}_{\rho_0=\ket{00}\bra{00}} = \braket{\hat{n}_2(\tau)}_{\rho_0=\ket{00}\bra{00}}$ and $\braket{\hat{n}_1(\tau)}_{\rho_0=\ket{10}\bra{10}} = \braket{\hat{n}_2(\tau)}_{\rho_0=\ket{01}\bra{01}}$. 
Considering this, it is clear that Eq.\ (\ref{Eq:g2tauPopulations}) is symmetric with respect to $\tau$. These equalities will not hold for $\gamma_o\neq 0$ and the asymmetry of $g^{(2)}_{12}(\tau)$ will be encoded in the coefficients $A_\pm$, $B_\pm$, $C_\pm$, $D_\pm$.

For balanced pumping ($\gamma_o = 0$), the second-order cross correlations can be expressed as
\begin{equation}
    g^{(2)}_{12}(\tau) = 1 - e^{-\gamma_e\vert\tau\vert}\sin^2(2\theta)\sin^2\left(\frac{\Delta E\tau}{2}\right),
\end{equation}
which we have written in terms of the mixing angle $\theta$ to clarify the manifestation of exciton delocalization in the second-order cross correlation for the case of balanced pumping. It is then simple to analyze the cross correlation for different regimes of exciton delocalization by varying $\theta$ while keeping $\Delta E$ fixed, as pointed out in Section \ref{Sec:DimerDescription}. In this case, we have $g^{(2)}_{12}(\tau)$ to be periodic with an angular frequency determined by the exciton energy gap $\Delta E$, and the amplitude of its oscillations to be determined by the mixing angle via the coefficient $\sin^2(2\theta)$. 
The amplitude is maximized for a completely delocalized state, and decreases as it becomes quasi-localized until finally the emissions become uncorrelated for all $\tau$ once the system is completely localized. 
In the case of balanced pumping we have Figs.\ \ref{Fig:g2tau_subplots}(a) and \ref{Fig:g2tau_subplots}(b) depicting $g_{12}^{(2)}(\tau)$ for mixing angles $\theta = \pi/8$ and $\theta = \pi/16$, respectively, which shows the decrease in amplitude. 
A mixing angle $\theta = 0$ corresponds to a completely localized state and results in emissions being uncorrelated at all times, $g_{12}^{(2)}(\tau) = 1$. 

Assuming once again that we have a pumping imbalance ($\gamma_o\neq0$) such that the second-order cross correlations are of the form in Eq.\ (\ref{Eq:g2tauelementary}), there will be an asymmetry due to the different initial conditions that are encoded in the constants $A_\pm$, $B_\pm$, $C_\pm$, $D_\pm$. 
It should be noted that the correlations are not periodic in this case because of the presence of $e^{\pm\frac{\delta}{\omega}\gamma_o\tau}$ terms, unless we consider the case of completely delocalized excitons as $\theta = \pi/4$ implies $\delta = 0$ which will reduce the exponential terms to constants. 
Figs.\ \ref{Fig:g2tau_subplots}(c) and \ref{Fig:g2tau_subplots}(d) show two examples of $g_{12}^{(2)}(\tau)$ when there is a pumping imbalance and $\delta\neq0$. 
In Figs.\ \ref{Fig:g2tau_subplots}(c) and \ref{Fig:g2tau_subplots}(d), the mixing angle is set to $\theta = \pi/8$ and $\theta = \pi/16$, respectively, and the cross correlations are neither symmetric nor periodic, although the oscillations are still characterized by $\omega$. 
The damping of the oscillations is more pronounced than that shown in Fig.\ \ref{Fig:g2tau_subplots}(c), but there were stronger correlations than those observed in Fig.\ \ref{Fig:g2tau_subplots}(b).

Note that the auto correlation for either emitter, $g_{11}(\tau)$ and $g_{22}(\tau)$, displays perfect anti-bunching behavior as each supports at most a single excitation. 
The second emission can occur only once the emitter is repopulated, either as a result of incoherent pumping or coherent energy transfer. 
They exhibit oscillatory behavior due to the coherent coupling, similar to that observed in the cross correlation. 
Their reliance on a repopulation process to produce an optical signal, however, means that the timescale for witnessing oscillatory behavior is reduced in comparison with cross correlations.
The cross correlation is therefore a stronger witness to the collective behavior within the dimer, as it is not restricted by the timescale of repopulation.

\begin{figure*}
    \centering
    \begin{minipage}{.5\linewidth}
        \includegraphics[width=\linewidth]{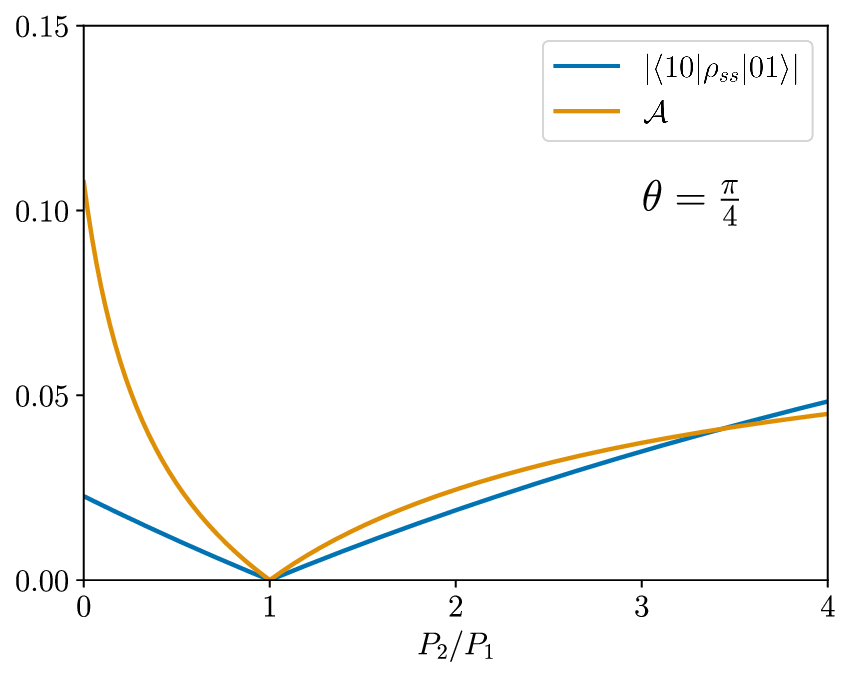}
    \end{minipage}%
    \begin{minipage}{.5\linewidth}
        \includegraphics[width=\linewidth]{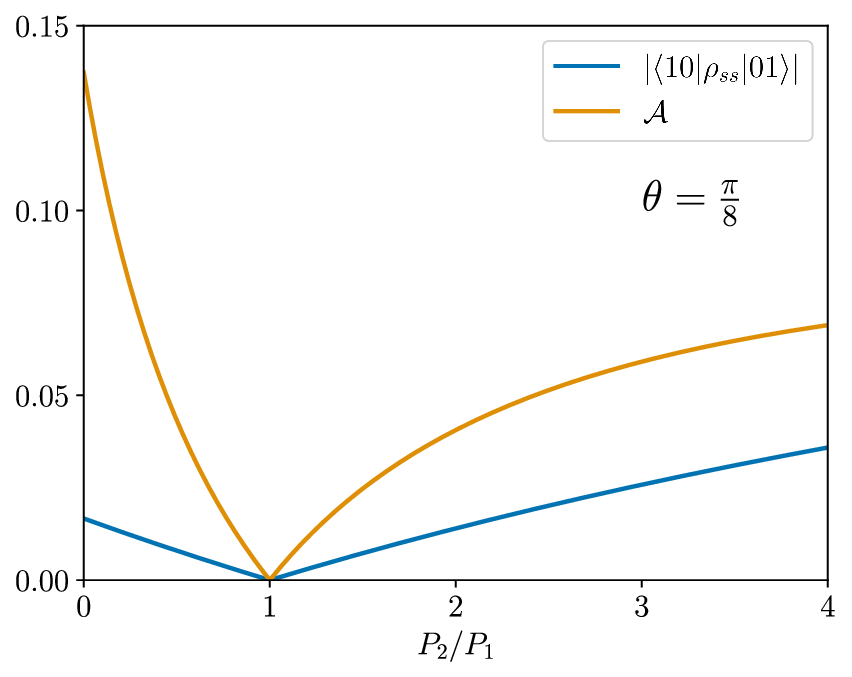}
    \end{minipage} 
    \caption{Cross correlation asymmetry quantified by Eq.\ (\ref{eq:correlation_asymmetry}) as a function of pumping imbalance for two values of the mixing angle $\theta$. 
    We also show the absolute value of the steady-state electronic coherence in order to demonstrate the association between the two. 
    Here we have set $\Delta E = 4\gamma$ and $P_1 = \gamma / 10$, similarly to Ref.\ \cite{SanchezMunoz2020May}. }
    \label{fig:g2tau_asymmtry}
\end{figure*}

We now consider the time asymmetry of the cross correlation function and highlight its association to coherence in the site basis, $\braket{10\vert \rho_{ss}\vert 01}$. 
To quantify the asymmetry, we introduce
\begin{equation}\label{eq:correlation_asymmetry}
    \mathcal{A} = \gamma\int_0^\infty\text{d}\tau \vert g_{12}^{(2)}(\tau) - g_{21}^{(2)}(\tau) \vert,
\end{equation}
as a unit-less figure of merit for the magnitude of cross correlation asymmetry. 
In Fig.\ \ref{fig:g2tau_asymmtry} we show both $\mathcal{A}$ and $\vert\braket{10\vert \rho_{ss}\vert 01}\vert$ as a function of the ratio $P_1 / P_2$ for two examples of the mixing angle $\theta$. 
We can see that when the imbalanced pumping is varied, the increase in site coherence is positively correlated with an increase in correlation asymmetry. 
When the incoherent pumping of the sites is balanced, we see that both quantities approach zero.
Therefore, the positive correlation between the two allows us to conclude that the observation of asymmetry in optical cross correlations is a witness to coherence within the heterodimer system.

\section{Conclusions}\label{sec:conclusions}
We have analyzed time-resolved second-order cross correlations for a prototypical acceptor-donor model consisting of two coherently interacting chromophores that are detuned from each other, and demonstrated their potential to exhibit clear signatures of both coherent energy transfer and exciton delocalization. 
We investigated the dynamical behavior of the cross correlation of the emissions from the steady-state for different pumping conditions. 
Under balanced pumping, the steady-state exhibits no coherence in the site basis, and joint emissions have an associated time-symmetric cross correlation whose oscillations characterize the timescale of coherent energy transfer, and whose deviations from unity depend strongly on the mixing angle $\theta$ and therefore may be used to quantify the degree of delocalization of the excitonic eigenstates. 
By contrast, under imbalanced pumping, the steady-state exhibits a site population difference proportional to the site basis coherences, and the cross correlations become time-asymmetric. 
The asymmetry of these cross correlations then report on the electronic coherence in the steady-state, as shown in Fig.\ \ref{fig:g2tau_asymmtry}. 
Therefore, we envision that time-resolved measurements of intensity cross correlations may inform on different quantum features such as exciton delocalization, coherent energy transfer, and steady-state coherence within light-harvesting complexes found in nature. 
We anticipate that the signatures we have identified may persist even under the influence of relevant non-Markovian and structured phonon environments, which are ubiquitous in biomolecular systems \cite{O'Reilly2014Jan,  Nation2024Feb}. 

\section*{Acknowledgments}
We gratefully acknowledge funding from the Gordon and Betty Moore Foundation (Grant GBMF8820).

\newpage
\widetext
\appendix

\section{Steady-state}\label{Appendix:SteadyStatePopulations}
The elements of the steady-state $\rho_{ss}$ are determined in the $\mathcal{B}$ basis by solving the matrix equation $L\Vec{p} = \Vec{0}$, where
\begin{equation}
    L = \begin{bmatrix}
        -P_1 - P_2 & \gamma_1 & \gamma_2 & 0 & 0 & 0 \\
        P_1 & -\gamma_1 - P_2 & 0 & \gamma_2 & iV & -iV \\
        P_2 & 0 & -\gamma_2 - P_1 & \gamma_1 & -iV & iV \\
        0 & P_2 & P_1 & -\gamma_1 - \gamma_2 & 0 & 0 \\
        0 & iV & -iV & 0 & -\gamma_e - i\delta & 0 \\
        0 & -iV & iV & 0 & 0 & -\gamma_e + i\delta
    \end{bmatrix}
\end{equation}
and $\Vec{p}$ is defined as
\begin{equation}
    \Vec{p} = \begin{bmatrix}
        \braket{00\vert\rho_{ss}\vert 00} \\
        \braket{10\vert\rho_{ss}\vert 10} \\
        \braket{01\vert\rho_{ss}\vert 01} \\
        \braket{11\vert\rho_{ss}\vert 11} \\
        \braket{10\vert\rho_{ss}\vert 01} \\
        \braket{01\vert\rho_{ss}\vert 10}
    \end{bmatrix},
\end{equation}
when imposing the condition $\text{Tr}(\rho_{ss}) = 1$. This is a reformulation of the quantum master equation $\partial_t\rho = 0$ in terms of the site states, where we omitted all uncoupled elements. Setting $\gamma_1 = \gamma_2 = \gamma$ and solving for $L\Vec{p} = \Vec{0}$ yields the steady-state populations given in the main text, Eqs.\ (\ref{Eq:SteadyStatePopulation1}-\ref{Eq:SteadyStatePopulation4}). 

\section{Populations and coherence dynamics for balanced pumping}\label{Appendix:DynamicsAppendix}
Here, we give the expressions for the populations and coherence dynamics in the case where both the radiative decay rates and the incoherent pumping rates for both chromophores are identical, that is, $\gamma_1 = \gamma_2 = \gamma$ and $P_1 = P_2 = P$ such that $\gamma_e = \gamma + P$ and $\gamma_o = 0$.

For an initial state $\rho_0 = \ket{00}\bra{00}$ we have
\begin{align}
    &\braket{\hat{n}_1(t)} = \frac{P}{\gamma_e}(1 - e^{-\gamma_e t}), \\
    &\braket{\hat{n}_2(t)} = \frac{P}{\gamma_e}(1 - e^{-\gamma_e t}), \\
    &\braket{[\sigma_1^+\sigma_2^-](t)} = 0.
\end{align}

For an initial state $\rho_0 = \ket{11}\bra{11}$ we have
\begin{align}
    &\braket{\hat{n}_1(t)} = \frac{1}{\gamma_e}(P + \gamma e^{-\gamma_e t}), \\
    &\braket{\hat{n}_2(t)} = \frac{1}{\gamma_e}(P + \gamma e^{-\gamma_e t}), \\
    &\braket{[\sigma_1^+\sigma_2^-](t)} = 0.
\end{align}

For an initial state $\rho_0 = \ket{01}\bra{01}$ we have
\begin{align}
    &\braket{\hat{n}_1(t)} = \frac{P}{\gamma_e} + \frac{e^{-\gamma_e t}}{\gamma_e\Delta E^2}\left(4V^2\left(\gamma\sin^2\left(\frac{\Delta Et}{2}\right) - P\cos^2\left(\frac{\Delta Et}{2}\right)\right) - P\delta^2\right), \\
    &\braket{\hat{n}_2(t)} = \frac{P}{\gamma_e} + \frac{e^{-\gamma_e t}}{\gamma_e\Delta E^2}\left(4V^2\left(\gamma\cos^2\left(\frac{\Delta Et}{2}\right) - P\sin^2\left(\frac{\Delta Et}{2}\right)\right) + \gamma\delta^2\right), \\
    &\braket{[\sigma_1^+\sigma_2^-](t)} = -\frac{Ve^{-\gamma_e t}}{\Delta E^2}\left(\delta(\cos(\Delta Et)-1) + i\Delta E\sin(\Delta Et)\right).
\end{align}

For an initial state $\rho_0 = \ket{10}\bra{10}$ we have
\begin{align}
    &\braket{\hat{n}_1(t)} = \frac{P}{\gamma_e} + \frac{e^{-\gamma_e t}}{\gamma_e\Delta E^2}\left(4V^2\left(\gamma\cos^2\left(\frac{\Delta Et}{2}\right) - P\sin^2\left(\frac{\Delta Et}{2}\right)\right) + \gamma\delta^2\right), \\
    &\braket{\hat{n}_2(t)} = \frac{P}{\gamma_e} + \frac{e^{-\gamma_e t}}{\gamma_e\Delta E^2}\left(4V^2\left(\gamma\sin^2\left(\frac{\Delta Et}{2}\right) - P\cos^2\left(\frac{\Delta Et}{2}\right)\right) - P\delta^2\right), \\
    &\braket{[\sigma_1^+\sigma_2^-](t)} = \frac{Ve^{-\gamma_e t}}{\Delta E^2}\left(\delta(\cos(\Delta Et)-1) + i\Delta E\sin(\Delta Et)\right).
\end{align}
Notice that $\braket{\hat{n}_2(t)}_{\rho_0 = \ket{01}\bra{01}} = \braket{\hat{n}_1(t)}_{\rho_0 = \ket{10}\bra{10}}$ which results in a symmetric $g^{(2)}(\tau)$.

\bibliographystyle{apsrev4-2}
\bibliography{mybib}

\end{document}